\documentclass[useAMS,usenatbib]{mn2e}
\usepackage{graphicx,amssym}
\newcommand{\bc}{\begin{center}}
\newcommand{\ec}{\end{center}}
\usepackage{color}
\newcommand{\comment}[1]{}
\title[Galaxy population in WDM and CDM cosmologies]
      {The galaxy population in cold and warm dark matter cosmologies}

\author[Wang et al. ]
       {Lan Wang$^{1}$$\thanks{Email:
           wanglan@bao.ac.cn}$, Violeta Gonzalez-Perez$^{2,3}$, Lizhi Xie$^{4}$, 
Andrew P. Cooper$^{2}$, \and Carlos S. Frenk$^{2}$,
Liang Gao$^{1}$, Wojciech A. Hellwing$^{3,5}$, John Helly$^{2}$, \and Mark R. Lovell$^{6,7}$,  
and Lilian Jiang$^{2}$
        \\      
        $^1$Key Laboratory for Computational Astrophysics, National
Astronomical Observatory, \\
Chinese Academy of Sciences, Datun Road 20A, Beijing 100012, China\\
	$^2$Institute for Computational Cosmology, 
Department of Physics, University of Durham, South Road, Durham DH1 3LE, UK\\
	$^3$Institute of Cosmology and Gravitation, 
University of Portsmouth,Burnaby Road, Portsmouth PO1 3FX, UK\\
        $^4$INAF - Astronomical Observatory of Trieste, via G.B. Tiepolo 11, I-34143 Trieste, Italy \\
        $^5$Janusz Gil Institute of Astronomy, University of Zielona G\'ora, ul. Szafrana 2, 65-516 Zielona G\'ora, Poland\\
        $^6$GRAPPA, Universiteit van Amsterdam, Science Park 904, NL-1098 XH Amsterdam, the Netherlands\\
        $^7$Instituut-Lorentz for Theoretical Physics, Niels Bohrweg 2, NL-2333 CA Leiden, the Netherlands}
 
\begin{document}

\date{Accepted 2016 ???? ??. 
      Received 2016 ???? ??; 
      in original form 2016 ???? ??}

\pagerange{\pageref{firstpage}--\pageref{lastpage}} 
\pubyear{2016}

\maketitle

\label{firstpage}

\begin{abstract}
We use a pair of high resolution N-body simulations implementing two 
dark matter models, namely the standard cold dark matter (CDM) 
cosmogony and a warm dark matter (WDM) alternative where the dark matter 
particle is a 1.5~keV thermal relic. We combine these simulations with the 
{\sc GALFORM} semi-analytical
galaxy formation model in order to explore differences between the resulting  
galaxy populations. We use {\sc GALFORM} model variants for CDM and WDM that 
result in the same $z=0$ galaxy stellar mass function by construction. We
find that most of the studied galaxy properties have the same values in these two models, 
indicating that both dark matter scenarios match current observational data 
equally well. Even in under-dense regions, where discrepancies in structure formation
between CDM and WDM are expected to be most pronounced, the 
galaxy properties are only slightly different. The only significant difference 
in the local universe we find
is in the galaxy populations of ``Local Volumes'', regions of radius $1$ to 8$\mathrm{Mpc}$ 
around simulated Milky Way analogues. In such regions our WDM model provides a better 
match to observed local galaxy number counts and is five times more likely than the CDM model 
to predict sub-regions within them that are as empty as the observed Local Void. Thus, 
a highly complete census of the Local Volume and future surveys of void regions could provide
constraints on the nature of dark matter.
\end{abstract}

\begin{keywords}
   (cosmology:) dark matter -- galaxies: formation -- galaxies: abundances -- galaxies: high-redshift
\end{keywords}

%%%%%%%%%%%%%%%%%%%%%%%%%%%%%%%%%%%%%%%%%
\section{Introduction}
\label{sec:intro}

%CDM has problems on small scales;

Most matter in the Universe is made of unknown exotic fundamental particles 
known as dark matter. Its existence has been inferred from
various observations, including galaxy rotational curves, gravitational
lensing, and the mass-to-light ratio of clusters \citep[see a recent 
review by][]{bertone2016}. The most popular model of
dark matter consists of a supersymmetric particle that has a
negligible velocity dispersion, allowing density perturbations
imprinted in the early Universe to persist down to very small
scales. This model is successful in matching a large body of
observations, from temperature fluctuations in the microwave
background \citep[e.g.][]{komatsu2011,planck2015} to the galaxy distribution today 
\citep[e.g.][]{frenk2012,sanchez2017}. Nevertheless, there has
been growing controversy about its validity on the scale of individual galaxies
and below. For example, it has been suggested that there are discrepancies 
between predictions 
and observations of the abundance of satellite galaxies around the Milky Way
\citep{klypin1999,moore1999}, of the kinetic stellar data of the Milky Way
satellites \citep{boylan2012}, and of the density profiles of low 
surface brightness dwarf galaxies \citep{moore1999,springel2008}. 

The most popular candidate for cold dark matter -- the hypothetical 
weakly interacting massive particle (WIMP) -- has not been detected despite a dedicated 
campaign of searches in colliders, underground laboratories \citep{akerib2014}, and 
the gamma-ray sky \citep[e.g.][]{ackermann2015}. It is therefore imperative to consider 
alternative dark matter candidates such as, for example, the {\rm keV}-scale gravitino 
or sterile neutrino. In addition to their specific particle physics identities, these 
two candidates differ from WIMPs in that they are warm dark matter (WDM) candidates 
rather than cold dark matter (CDM) ones.

The WDM model has been considered to be
one possibility for solving the issues facing the CDM cosmology on small
scales. In the past decade, many papers have modelled
WDM structure formation \citep[e.g][]{bode2001,knebe2002, knebe2003,
busha2007, colin2008, zavala2009, smith2011,schneider2012, lovell2012, destri2013,
angulo2013, benson2013, kamada2013, lovell2014, bose2016a,ludlow2016,liran2016,liran2016b}. 
In the WDM model, the dark matter particles have intrinsic thermal velocities,
and these velocities influence the small scale structure formation of the Universe
mainly in two different ways. Firstly,  the motion of warm dark matter
particles would quench the growth of structure below some free-streaming scale
(the distance over which a typical WDM particle travels). Since
small and dense haloes do not form below the free-streaming-scale, the
dark matter haloes that surround galaxies in a WDM model have far less
substructure compared to their CDM counterparts,
which may help alleviate the satellite abundance
problem \citep{bode2001}. Secondly, according to the phase-space density theory 
\citep{tremaine1979}, the primordial velocities of collisionless WDM particles 
impose a finite phase-space density, which ultimately prevents the formation of a 
cuspy profile in WDM haloes \citep{shao2013}. Consequently, the innermost density 
profiles of WDM haloes are predicted to be cored instead of the cuspy ones 
predicted by CDM simulations. Nevertheless, recent progress on
understanding the density profile of WDM haloes indicates that
a realistic warm dark model cannot account for the sizes of cores in the density profiles
of dwarf galaxies inferred from observations \citep{shao2013,maccio2013}.
The observed sizes of cores of dwarf galaxies
\citep{gilmore2007,walker2011} should be produced by some 
baryonic physics, such as outflows \citep{navarro1996}, supernovae feedback 
and baryon clumps \citep{delpopolo2016}, or reflect a 
different dark matter candidate such as self-interacting dark matter
\citep{vogelsberger2012, zavala2013}.

%studies on WDM galaxy formation.
To explore baryonic physics in WDM cosmogonies, studies of galaxy formation 
have been carried out by using either semi-analytic
models \citep{menci2012, menci2013, nierenberg2013, kang2013,
  lovell2015, bose2016,lovell2016b}, or hydrodynamical simulations 
\citep{herpich2014, maio2015, colin2015, power2016,lovell2016c}. 
These works have studied observables including global galaxy properties like
luminosity/stellar mass functions and galaxy colour/star formation 
\citep{menci2012, menci2013, kang2013,herpich2014}, the inner structure of galaxies 
\citep{herpich2014,colin2015,power2016, gonzalez-samaniego2016}, and 
the properties of satellite galaxies around galaxies comparable to the Milky Way 
\citep{nierenberg2013, lovell2015, lovell2016b}. The masses of the 
WDM particles adopted in these studies vary from $m_x=0.5{\rm kev}$ to $3.3{\rm kev}$.
These values were inferred from constraints provided by different
observations: the numbers of Milky Way satellites \citep{polisensky2010}, X-ray
observations of the Andromeda galaxy \citep{watson2012}, galaxy counts
at high redshift \citep{pacucci2013}, high redshift long $\gamma$-ray
bursts \citep{deSouza2013} and the high redshift Ly-$\alpha$ forest
data \citep{viel2013}.

%The mass resolutions of previous WDM simulations are
%usually down to a few times $10^7M_{\odot}$, except for the cases of
%very small simulated boxes  \citet[e.g., $10h^{-1}Mpc$ in
%][]{kamada2013} or only analysing a few (galactic) haloes or
%filaments \citep{knebe2003,lovell2012,reed2015,ludlow2016} that focus
%on subhalo properties or local group.

In this work we present two simulations, one in a CDM and the other in a WDM 
cosmogony. The two simulations have identical initial conditions, except for a
truncation on small scales in the power spectrum of the WDM simulation. Our simulations 
are state-of-the-art both in the volume of the universe they simulate and their mass 
resolution. We exploit these advantages by combining the two simulations 
with {\sc GALFORM}, a semi-analytic model of galaxy formation. This allows 
us to explore systematic differences between the galaxy populations formed 
in the two cosmologies. Historically, cosmological WDM simulations,
and hence semi-analytic galaxy formation models built on them,
have suffered from uncertainties introduced by spurious self-bound DM clumps 
arising from numerical discreetness effects 
\citep{wangjie2007}. These spurious haloes have been carefully removed from the  
halo catalogue on which our GALFORM model is based by applying the procedure of 
\citet{lovell2014}, which improves the reliability of comparisons between our 
results and observations. 
We note that hydrodynamical simulations by \citet{gao2007} and \citet{gao2015} have
demonstrated a novel star formation mechanism in WDM, which occurs in
filaments rather than in collapsed dark matter haloes. Estimates of the star 
formation efficiency of this process are currently highly uncertain, however, 
so we neglect it in this study. 

The outline of our paper is as follows. In Section~2 we introduce our
simulation sets and the semi-analytic models used to
populate dark matter haloes with galaxies. In Section~3, we study the statistics 
of galaxy properties 
at both low and high redshift. In Section~4, we compare the
properties of voids and their constituent galaxies, because the differences between 
WDM and CDM cosmogonies 
are expected to be most pronounced in these low density regions. 
Finally, we present a discussion and our conclusions in Section~5.

%%%%%%%%%%%%%%%%%%%%%%%%%%%%%%%%%%%%%%%%%
\section{simulations and galaxy formation models}
\label{sec:sim}

\subsection{Simulations}

\begin{table} 
  \caption{The details of the two simulations 
used in this study. $L$ is the side length of the simulation box; $N$ is the 
total number of particles; $m_{\rm dm}$ is the particle mass; $\epsilon$ is the force 
softening length; $m_{WDM}$ is the assumed mass of the thermal WDM particle in 
the WDM simulation. Cosmological parameters are consistent with the WMAP7 
results \citep{komatsu2011}.
$\Omega_{\rm m}$, $\Omega_\Lambda$, and $\Omega_{\rm b}$ are respectively 
the cosmological average densities of matter, dark energy and baryonic matter in units 
of the critical density at redshift zero. $H_0$ is the Hubble parameter.
$\sigma_8$ is the square root of the linear variance of matter 
distribution when smoothed with a top-hat filter of radius $8~h^{-1}$Mpc.
  }
\begin{center}
\begin{tabular}{|l|r|}
\hline
Property &   value  \phantom{000} \\
\hline
   $L$ (comoving $h^{-1}Mpc$)             &   70.4\phantom{00000}  \\
   $N$             &    $1620^3$\phantom{0000} \\
   $m_{\rm dm}$ ($h^{-1}M_{\odot}$)    &    $  6.20\times10^{6}$\phantom{} \\
   $\epsilon$~(comoving $h^{-1}$kpc)  &    1.0\phantom{000000}   \\
   $m_{\rm WDM}$(${\rm kev}$, for the WDM simulation)  &    1.5\phantom{000000}   \\
\hline
   $\Omega_{\rm m}$             &    0.272\phantom{0000} \\
   $\Omega_\Lambda$             &    0.728\phantom{0000} \\
   $\Omega_{\rm b}$             &    0.04455\phantom{00} \\ 
   $h \equiv H_0$/(100 km\,s$^{-1}$\,Mpc$^{-1})$ &  0.704\phantom{0000} \\
   $\sigma_8$                  &    0.81\phantom{00000} \\
   $ n $                 &    0.967\phantom{0000} \\
   %$Y$                         &    0.248\phantom{00}\\
\hline
\end{tabular}
\end{center}
\label{tbl:cosmo}
\end{table}

The numerical simulations used in this study comprise a pair of high
resolution dark matter-only simulations. The two simulations are
identical except for the different nature of dark matter, one with
standard CDM and one with a WDM
model. The former one has been introduced in detail by \citet{hellwing2016}.
For each of the simulations  $1620^3$ particles are evolved
within a box of length $100$~Mpc on a side. Thus, the individual
particle mass is $6.2 \times 10^6 {\rm M_{\odot}}$. Cosmological parameters 
consistent with the WMAP7 results \citep{komatsu2011} are adopted. The details of the
simulation setup and cosmological parameters are presented in Table~\ref{tbl:cosmo}.

The initial conditions for the simulations are generated at redshift
$z=127$, made by a $3072^3$ Fourier grids with initial phases taken
from the multiscale Gaussian field called Panphasia \citep[see ][for details]{jenkins2013}. 
The transfer function of the CDM run is computed
with the Boltzmann code CMBFAST \citep{seljak1996}. For the WDM run,
the linear power spectrum is calculated by sharply truncating \citep{viel2005}
the CDM power below a free-streaming scale corresponding to that of
$1.5$~{\rm keV} relic particle. This choice of WDM particle mass is warmer than 
the latest Lyman-$\alpha$ forest constraint allows \citep{viel2013, garzilli2015}. 
This model is used because it is extreme, and therefore
emphasises the differences between CDM and WDM cosmologies.
Note that the two simulations use the
same random phase to initialize the Gaussian fields, allowing a
straightforward comparison between the two dark matter models.

\begin{figure*}
\bc
\hspace{-0.4cm}
\resizebox{16cm}{!}{\includegraphics{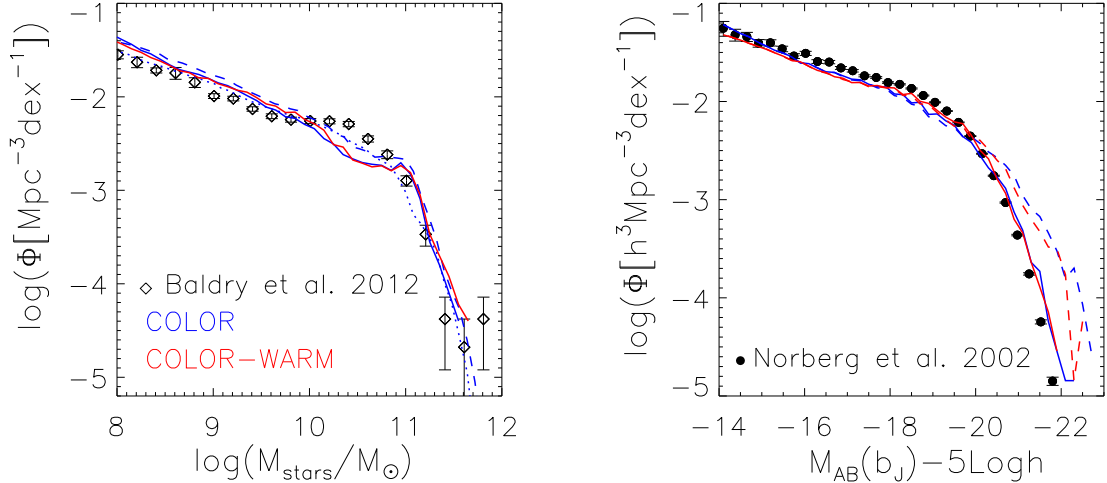}}\\%
\caption{ Left panel: stellar mass functions
of galaxies in the two models. Black symbols are the observations 
given by \citet{baldry2012}, corrected to a Kennicutt IMF \citep{lacey2016}. 
The blue solid line is for the COLOR model, 
and the red solid line is for 
 COLOR-WARM. The blue dashed line is the result of the GP14 
model combined with MS-W7 simulation as shown in the GP14 paper. 
The blue dotted line shows the stellar mass function of the GP14 model 
combined with MS-W7 simulation obtained from the model broad band photometry 
by SED-fitting as given in the GP14 paper (see text for more details).
Right panel: $b_J$ band galaxy luminosity function in the two models. 
Solid lines are results including the attenuation by dust while dashed lines are 
the ones without it. Black dots with error bars are 2dF 
results of \citet{norberg2002}.
 }
\label{fig:SMF}
\ec
\end{figure*}

The simulations are performed with the GADGET-3 Tree-PM N-body code, which
is an updated version of the public GADGET-2 code \citep{springel2005}. For 
each simulation, we have record  $80$ snapshots roughly logarithmically
spaced in redshift between $z=40$ and $z=0$.

For each output snapshot, dark matter haloes are identified using a
Friends-Of-Friends (FOF) algorithm \citep{davis1985} with a linking
length of $b=0.2$ in units of the mean inter-particle separation. Then
the SUBFIND algorithm \citep{springel2001a} is applied to identify
self-bound and locally over-dense substructures within each FOF
halo. Substructures with more than 20 particles are considered as
subhaloes. Halo/subhalo merger trees are constructed following the
method described by \citet{jiang2013}.

%This tree building method looks several snapshots ahead to link
%progenitor and descendant subhaloes robustly, and makes sure that the
%halo formation histories are strictly hierarchical.

%For the warm dark matter simulation, the warm dark matter particle is 
%chosen to have thermal mass of $m_x=1.5{\rm kev}$. The free streaming mass 
%below which scale small structures are smeared
%out due to thermal motion of particles, as given by \citet{bode2001} is
%$2.924\times10^{9}\,h^{-1}{\rm M}_{\odot}$.

WDM simulations are affected by numerical
discreetness, which causes artificial fragmentation of the smooth
filaments. This results in an effective halo mass
resolution limit \citep{wangjie2007} of $2.42\times10^{9}\,h^{-1}{\rm
M}_{\odot}$ in our WDM simulation, below which most structures identified by 
our halo finder are spurious (see Appendix A and Fig.~\ref{fig:haloMF}). Some of
the spurious haloes are massive enough that gas condenses and forms stars within them, which 
would affect the predictions of semi-analytic modelling even above this mass limit. 
We therefore follow the method described in \citet{lovell2014} to carefully 
identify and remove these spurious haloes from the halo merger trees. 

Hereafter we refer to the simulation with CDM as COLOR
(COco LOw Resolution simulation). 
{\footnote {This name was applied by
\citet{hellwing2016} to the simulation labelled ``DOVE'' in
\citet{jenkins2013, sawala2015,fattahi2016}.}} We refer to the simulation with
WDM of mass $m_x=1.5{\rm kev}$ as
COLOR-WARM \citep[this is identical to the simulation ``COLOR-1.5'' in][]{ludlow2016}.

\subsection{Model galaxies}

The semi-analytic galaxy formation model used in this study is the {\sc galform} 
model described by \citet{gonzalez2014} (hereafter GP14\footnote {This model is 
available through the Millennium Data Base:
http://virgodb.dur.ac.uk.}), which builds on earlier work by  \citet{cole2000,bower2006} 
and \citet{lagos2012}. The GP14 model was
calibrated on the MS-W7 simulation \citep{guo2013,lacey2015}, which
adopts the same WMAP7 cosmology as the simulations used in this work 
but has a much larger volume 
$(500h^{-1}Mpc)^3$ and a lower mass resolution of
$9.35\times10^8h^{-1}M_{\odot}$. The GP14 model successfully reproduces a 
wide range of observations \citep{gonzalez2014,lagos2015, merson2016}.

The physical and numerical parameters of a semi-analytical model such as
GP14 are typically calibrated with reference to a particular set of 
simulation merger trees. The particle mass resolution, the time resolution, 
and (more physically) the mass assembly histories of
DM haloes in a particular cosmogony all influence the predicted properties of 
galaxies to some degree \citep{lee2014}.
{\sc GALFORM} includes numerical prescriptions intended to ensure that the properties 
of galaxies well-resolved in the simulation used for calibration are converged with 
respect to further increases in mass and time resolution. We nevertheless find that 
direct application of the original GP14 model to the COLOR 
simulation results in global statistics for the galaxy population that are 
slightly different from those obtained in the MS-W7 calibration (for example the 
stellar mass function, see Fig.~\ref{fig:SMF}).  We attribute this primarily to 
sample variance (COLOR has a volume $358$ times smaller than that of MS-W7); 
this issue is addressed in detail in
Appendix~\ref{app:resolution}.

In order to mitigate effects of these differences between 
CDM and WDM galaxy properties, we recalibrate the parameters of the model 
we apply to the COLOR-WARM simulation such that they yield the same $z=0$ galaxy 
stellar mass function as the original GP14 model applied to COLOR (in the mass 
range $10^9<M_{\rm stars}/(h^{-1}M_{\odot})<10^{11}$). We refer to this WDM-based 
recalibration of the model as Re-GP14. This recalibration introduces minimal 
changes to parameters controlling the strength of supernovae feedback (relatively 
weaker than GP14 in low-mass galaxies) and AGN feedback -- full details are given 
in Appendix~\ref{app:regp14}. We stress that we have \textit{not} altered the 
parameters of the original GP14 model that we apply to COLOR. Consequently, our 
comparison is built on two models that make essentially identical predictions for 
the distribution of galaxy masses in the simulation box, but which employ slightly 
different models of baryonic physics to achieve this (due to effects 
of differences in structure formation between CDM and WDM).

On this basis, in the following sections, we explore how other galaxy 
properties differ between the two models: the original
GP14 model applied to the COLOR simulation, and the Re-GP14 model applied
to the COLOR-WARM simulation. For simplicity, when comparing 
results, we refer to COLOR $+$ GP14 model results as COLOR, and
COLOR-WARM $+$ Re-GP14 results as COLOR-WARM.

%%%%%%%%%%%%%%%%%%%%%%%%%%%%%%%%%%%%%%%%%
%\section{comparison of galaxy properties}

\section{Global galaxy properties in the two dark matter models}
\label{sec:galaxies}

In this section we compare the statistics of the galaxy populations in the two
models, both in the local universe and at high redshift.  We begin 
with a comparison of each model to observed stellar
mass and luminosity functions in the local
universe and the two point galaxy correlation function
as a function of stellar mass. These statistics
provide the most fundamental description of how galaxies populate dark matter 
haloes in each model. We then proceed to compare their stellar mass functions 
at high redshifts and the evolution of the star formation rate density of the universe.
Note that normally the plausible semi-analytic models are matched with observations
at $z=0$ and can be accepted within a larger variation than the errors of the 
observation data \citep{bower2010,benson2014,rodrigues2016}, and the two model results 
can be considered to be similar within that variation.

\begin{figure*}
\bc
\hspace{-0.4cm}
\resizebox{15cm}{!}{\includegraphics{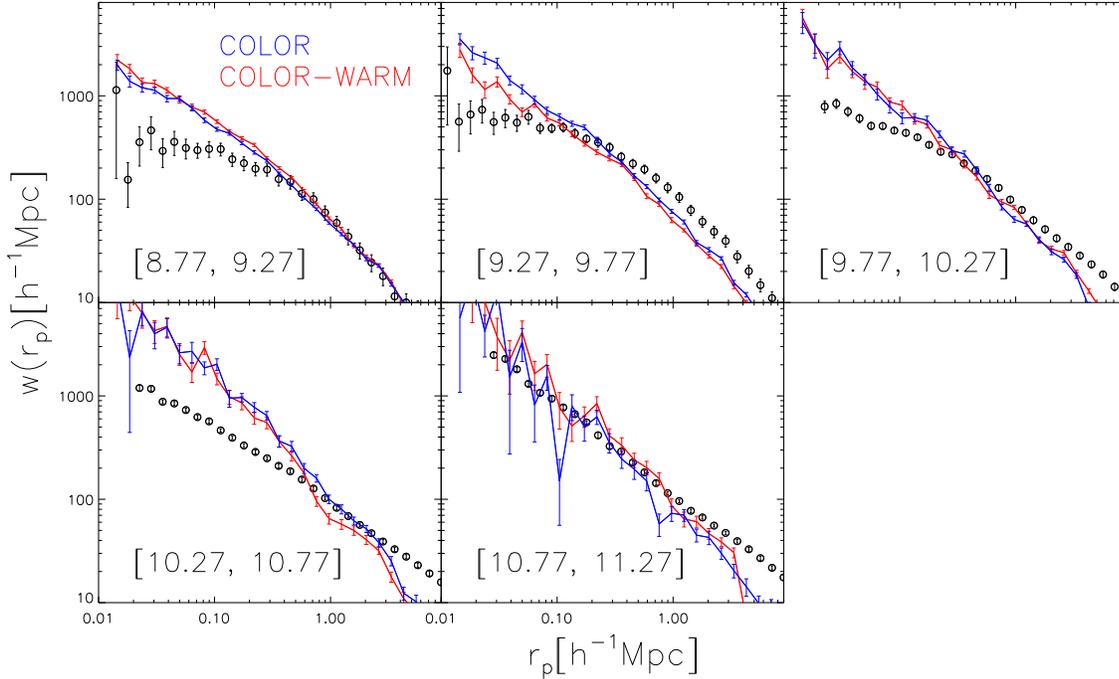}}\\%
\caption{ 
Projected two-point correlation functions of galaxies in the two models
binned in stellar mass. Error bars are from bootstrapping simulation data \citep{barrow1984}.
The circles with error bars show the SDSS DR7 results  \citep{li2006,guo2011}. 
}
\label{fig:CF}
\ec
\end{figure*}

\subsection{Present day stellar mass and luminosity functions}
The left panel of Fig.~\ref{fig:SMF} presents the $z=0$ galaxy stellar 
mass functions of the CDM and WDM runs. The blue solid line corresponds 
to the COLOR result and the red solid line to the result of 
COLOR-WARM. As described in the section
2.2, we calibrate the stellar mass function of COLOR-WARM to reproduce 
that of COLOR at $z=0$ over the stellar mass range
$10^9<M_{stars}/(h^{-1}M_{\odot})<10^{11}$.  The right-hand panel of Fig.~\ref{fig:SMF} 
shows the  $b_\mathrm{J}$-band luminosity function. 
The close correspondence between the $z=0$ luminosity functions  of COLOR-WARM and 
COLOR reflects the stellar mass function calibration.

In the left panel of Fig.~\ref{fig:SMF}, the blue dashed line is 
the result of the GP14 model applied to the MS-W7 simulation. The difference 
between the blue solid and dashed lines is mainly due to
cosmic variance (see Appendix~\ref{app:resolution} for further details). The blue 
dotted line in the same panel shows another version of this GP14 $+$ MS-W7 mass 
function, in which the stellar masses are obtained by spectral energy
distribution (SED) fitting, following an algorithm similar that used in observations
\citep{baldry2012,mitchell2013, gonzalez2014}. This reduces the inferred
abundance at both low and high stellar masses, and smooths the bump near
the ``knee'' of the function. We show it here to emphasize that discrepancies
between our baseline GP14 model and observations
can be reduced when model stellar masses are obtained
from SED fitting. This more complex calculation of the mass function 
is not used in the results we present here for COLOR and COLOR-WARM though, because our 
focus is on the difference between CDM and WDM models, rather than between 
either model and observations. 

\subsection{Galaxy correlation function}
The two-point correlation function is a basic measure
of the spatial clustering of galaxies. The dependence of galaxy
clustering as a function of various intrinsic galaxy properties has been well determined by 
observations\citep[e.g,][]{norberg2002,li2006,farrow2015,shi2016}. 
However, in galaxy formation models it is still difficult to
reproduce the amplitude of correlation functions for low mass galaxies
and on small scales \citep{guo2011, wang2012, campbell2015}. 
%Besides, degeneracy exists between cosmology and baryonic recipes 
%in shaping galaxy correlation functions\citep{wang2008}.

In Fig.~\ref{fig:CF}, we compare the projected galaxy correlation functions in 
different stellar mass bins from our simulated CDM and WDM galaxy
populations. Due to the limited box size of the simulations, these correlation 
functions fall off more rapidly than the data at large scales \citep{orsi2008,campbell2015}. 
For the box size of our simulations, the correlation functions can be measured 
only up to scales of a few megaparsecs. 
The correlation functions of the CDM and WDM model variants
are in a good agreement for all stellar masses. Some difference is seen for the stellar mass
bin $M_{stars}=10^{9.27-9.77}M_{\odot}$, but with insufficient statistics
 to draw firm conclusions.
%galaxies in COLOR-WARM cluster more weakly than those in COLOR at small scales. 
%This is due to the fact that, in this stellar mass bin, \adr{ galaxies in
%COLOR are hosted by dark matter haloes more massive by $\sim 33$ percent in the median
%than galaxies in the equivalent bin for COLOR-WARM. }\adb{many of the galaxies in COLOR are in haloes with mass greater than $10^{12}M_{\odot}$, whose clustering becomes much stronger than haloes less massive \citep{gao2005}. See the stellar mass-- halo mass relations shown in Fig. B2.} At other stellar masses, galaxies have more
%similar distributions of host halo mass. 

\subsection{Stellar mass functions at high redshifts}

\begin{figure*}
\bc
\hspace{-0.4cm}
\resizebox{18cm}{!}{\includegraphics{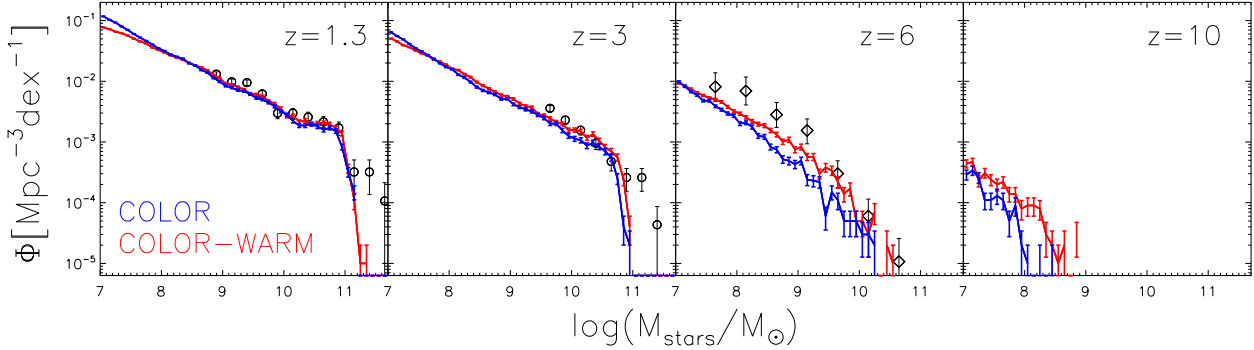}}\\%
\caption{ 
Galaxy stellar mass functions at redshifts of 1.3, 3, 6 and 10 in 
the two models. COLOR results are shown in blue and COLOR-WARM results in
red, with Poisson error bars. Black circles with error bars in the two leftmost 
panels are the observational complete data results from \citet{kajisawa2009} 
in redshift intervals of [1.0, 1.5] and [2.5, 3.5], respectively. 
Diamonds are observational data from \citet{gonzalez2011} at $<z>=5.9$. Observed stellar 
masses are adjusted to the assumption of a Kennicutt IMF following \citet{lacey2016}. 
}
\label{fig:highzSMF}
\ec
\end{figure*}

Although the stellar mass functions of our GP14 and Re-GP14 models are almost 
identical by construction at $z=0$, they are free to differ at higher redshifts. 
Such differences are expected from the different parameter choices that 
Re-GP14 requires to satisfy the observed constraint at $z=0$. 
These parameter differences most likely reflect fundamental differences in how 
structure formation proceeds in CDM and WDM  \citep{bode2001,knebe2002, angulo2013,gao2015}.

In Fig.~\ref{fig:highzSMF} we plot stellar mass functions at four redshifts in 
both dark matter models. At $z<3$, the stellar mass functions 
are quite similar. Towards higher redshift, however, massive galaxies are more 
abundant in COLOR-WARM than in COLOR. At $z=6$, the difference is about 3 times
at most, and at $z=10$, the difference can be as high as 9 times.
This is consistent with the results of 
\citet{bose2016}, who found that, with the same semi-analytic model applied to a 
cold and a $3.3$kev warm dark matter simulation, at redshifts greater than 5, 
the amplitude of the UV luminosity function in a WDM model was higher than in a 
CDM model \citep[although see also][who do not find an excess 
of bright galaxies at high redshift in a 1.5 kev WDM model]{dayal2015}. 
\citet{bose2016} attributed this to the fact that, in their models 
the brightest galaxies form  through  merger-triggered starbursts at high redshift, 
and this mechanism is more 
efficient in the WDM cosmology. We have checked that in our models, at $z>3$, 
there are indeed more starburst galaxies in COLOR-WARM than in COLOR, consistent
with this explanation.
%Bose et al. proposed that, at high redshift, the brightest galaxies form  through  merger-triggered starbursts of the first generation galaxies. in a WDM cosmogony, the first generation galaxies that form are more massive and more gas-rich than their CDM counterparts, whose cold gas is limited by supernovae feedback.  

%\adg{In addition to the explanation proposed by \citet{bose2016}, we note that the 
%excess of massive galaxies in the WDM model may be partially due to the fact 
%that, at high redshift there are more massive dark matter haloes in WDM than 
%in CDM cosmology. At $z=6.1$ the excess in the halo mass function
%is about 0.1-0.2 dex for haloes more massive than $\sim 10^{11}M_{\odot}$ (see Fig.~\ref{fig:haloMF}). 
%The weaker supernovae feedback in our Re-GP14 model leads to more efficient star 
%formation in these haloes, giving rise to an even greater difference in the stellar 
%mass function.}

\subsection{Cosmic star formation rate densities}

\begin{figure}
\bc
\hspace{-0.4cm}
\resizebox{8cm}{!}{\includegraphics{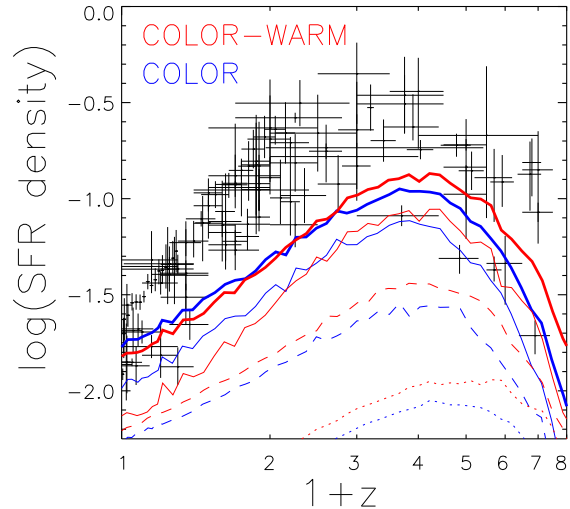}}\\%
\caption{Cosmic star formation rate density $\rho_{\mathrm{SFR}}$ (in units of 
$h^{2}\mathrm{M_{\odot}yr^{-1}Mpc^{-3}}$) as a function of redshift in the models. 
COLOR results are shown as blue lines and COLOR-WARM results as red lines.
Thick solid lines show the total SFR density. Thin lines with different line 
styles show the contributions to the total density from galaxies of different 
stellar mass ranges \textit{at the redshift of observation} 
as follows. Thin solid: the top 3~per~cent massive galaxies; dashed: galaxies with
stellar mass between the 85~per~cent and 97~per~cent distributions; 
dotted: the 85~per~cent least massive galaxies. 
As reference, black crosses are observational estimates compiled by \citet{hopkins2007}, 
%The grey shaded region shows the 1-$\sigma$ confidence interval of the compilation of \citet{wilkins2008}. Gold symbols  are the results of \citet{bouwens2009}, including contributions from highly dust obscured galaxies and ULIRGs. 
where SFRs have been 
adjusted to the assumption of a Kennicutt IMF following \citet{lacey2016}.}
\label{fig:SFRD}
\ec
\end{figure}

As discussed in the previous subsection, massive galaxies at high redshift form 
more efficiently in WDM than in CDM models. This effect is also apparent in the 
evolution of the cosmic star formation rate (SFR) density. In Fig.~\ref{fig:SFRD}, 
the thick solid blue line shows the total SFR density in COLOR, and the thick 
solid red line its equivalent in COLOR-WARM. All galaxies with stellar mass more massive than
$10^7M_{\odot}$ are included in the calculation. At $z<3$, the total SFR densities 
in the two models differ by less than 0.07~dex. At higher redshift, the total 
SFR density in COLOR-WARM is larger than that in COLOR, up to  $\sim0.3$~dex. 

We also plot in Fig.~\ref{fig:SFRD} the contributions to the total SFR density 
made by galaxies in different ranges of stellar mass. At each redshift, we
calculate the 85~per~cent and 97~per~cent distributions of galaxy stellar mass
\footnote{At z=0, the 85~per~cent and 97~per~cent distributions of galaxy
stellar masses for COLOR are $10^{8.63}$ and $10^{9.84}M_{\odot}$,
while those for COLOR-WARM are $10^{8.87}$ and $10^{10.01}M_{\odot}$.}. In Fig.~\ref{fig:SFRD},
the contributions of the most massive 3~per~cent galaxies as a function of redshift
are plotted in thin-solid lines. Dashed lines give contributions of galaxies with
stellar mass between the 85~per~cent and 97~per~cent distributions, and dotted lines
are the contributions of the least massive 85~per~cent galaxies.  
At all redshifts, the 
total SFR density is dominated by the most massive galaxy populations at the 
time. Comparing the two dark matter models, we see that for $z> 2$ the SFR 
densities of the most massive galaxies from COLOR-WARM are always higher 
than those in COLOR. 
The difference can be as high as 0.2 dex, corresponding to the difference in 
the total SFR density. This again reflects the more efficient formation of 
massive galaxies at redshifts $z> 2$ in the WDM model, consistent with the
picture described in section 3.3.

\begin{figure*}
\bc
\hspace{-0.4cm}
\resizebox{12cm}{!}{\includegraphics{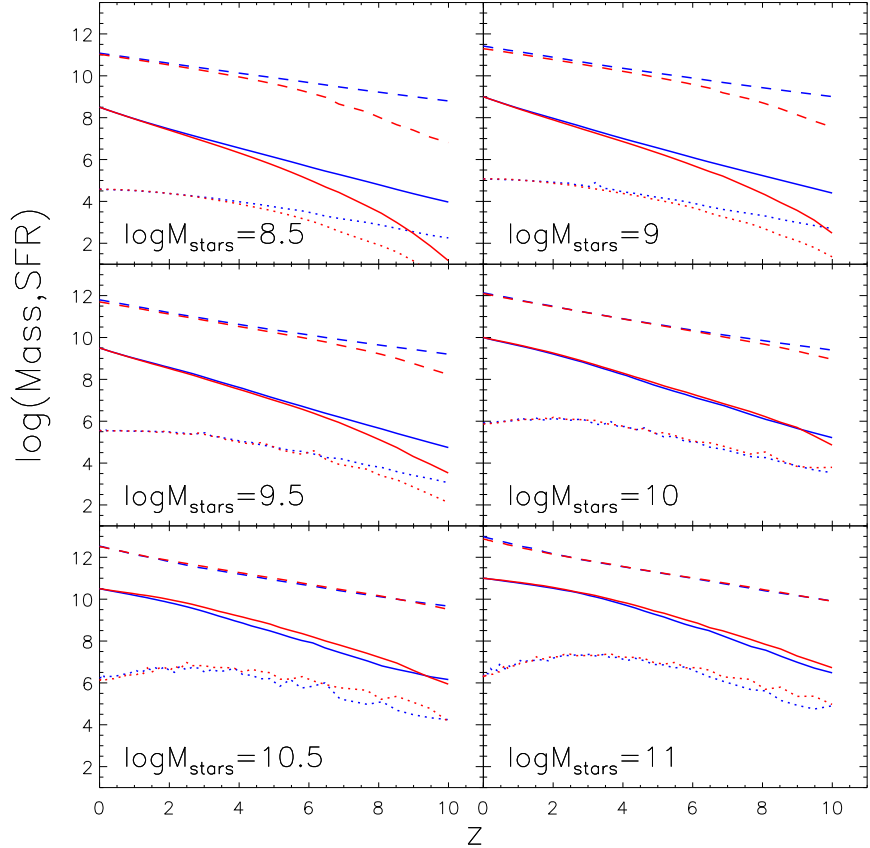}}\\%
\caption{Mean stellar mass (solid lines) and halo mass (dashed lines) 
assembly histories of central galaxies as a function of 
redshift at six stellar masses at z=0 (with bin width 0.2 in logarithm). 
Masses are in units of solar mass. Dotted lines 
show the SFR histories of these galaxies, in unit of $h^{-1}M_{\odot}/Gyr$. For clarity,
dotted lines are moved downward by 3 dex. Blue lines show 
COLOR results and red lines are from COLOR-WARM.
 }
\label{fig:checkhis}
\ec
\end{figure*}

\subsection{Mass assembly and SFR histories}
We can use the galaxy merger trees associated with our models to trace the 
stellar mass and halo mass of $z=0$ galaxies backwards in time. In 
Fig.~\ref{fig:checkhis}, dashed lines show the mean halo mass growth history 
for the main progenitor\footnote{We define the main progenitor by identifying 
the most bound ``core'' of dark matter particles, rather than the most 
massive one, following the method of \citet{jiang2013}. } of
central galaxies at $z=0$, split into six logarithmic bins of $z=0$ 
stellar mass bins (each of width $0.2$~dex). Solid lines denote the corresponding mean growth 
histories of stellar mass. Dotted lines are the corresponding mean SFR histories.
%\adb{(the first derivatives of the stellar mass history)}. 
As in previous figures, COLOR results are shown by blue lines, COLOR-WARM results by red lines.

Fig.~\ref{fig:checkhis} clearly shows that galaxies less massive than 
$10^{10}M_{\odot}$ at $z=0$ in COLOR-WARM assemble both their halo mass and
stellar mass later than in COLOR, with lower SFR at high
redshifts. The differences between the two models are larger for lower
mass galaxies. For galaxies with stellar mass of $10^{8.5}M_{\odot}$ at $z=0$, 
their mean stellar masses differ by more than 100 times at $z>9$, and the mean 
halo masses differ by more than 30 times.
For more massive galaxies, at a fixed stellar mass, halo mass 
grows at a similar rate in the two models but stellar mass assembles earlier (and SFR
is correspondingly higher) at high redshift in COLOR-WARM. The difference
in stellar mass is less than 3 times up to $z=10$. 

For galaxies in the range $10^{10} < M_{\star} < 10^{10.5}\,M_{\odot}$, at very high
redshift ($z>9$) there is a trend such that stellar masses in the
WDM model are lower than those in the CDM model. This is due to the later 
formation of the earliest progenitors in the WDM model. At redshifts lower 
than $\sim 9$, galaxies in the WDM model catch up 
through more gas-rich mergers, and have higher stellar 
mass in WDM than in CDM. These differences are consistent with those reported by
\citet{bose2016}, who used initial power spectrum corresponding to $3.3$~keV
thermal relic mass particle..

%void galaxy comparison
%%%%%%%%%%%%%%%%%%%%%%%%%%%%%%%%%%%%%%

\section{Galaxy populations in voids and in the Local Volume}

Numerical simulations of structure formation indicate that the
differences between WDM and CDM models are most pronounced on small scales
and in under-dense regions \citep[e.g.,][]{bode2001,knebe2002,
  knebe2003, angulo2013}. Anticipating that these differences may in turn influence the 
properties of galaxies in such regions, we explore voids in the distribution
of galaxies, and analogues of the Local 
Volume and the Local Void in our models. In the nearby Universe,  
galaxy counts are roughly 80~per~cent complete down to a faint limit of 
$m_{B}=17.5$ \citep{karachentsev2014}, which makes it possible to compare faint
galaxies with the observational data in order to constrain the
identity of dark matter.

\subsection{Galaxy population in Voids}

\begin{figure*}
\bc
\hspace{-0.4cm}
\resizebox{17cm}{!}{\includegraphics{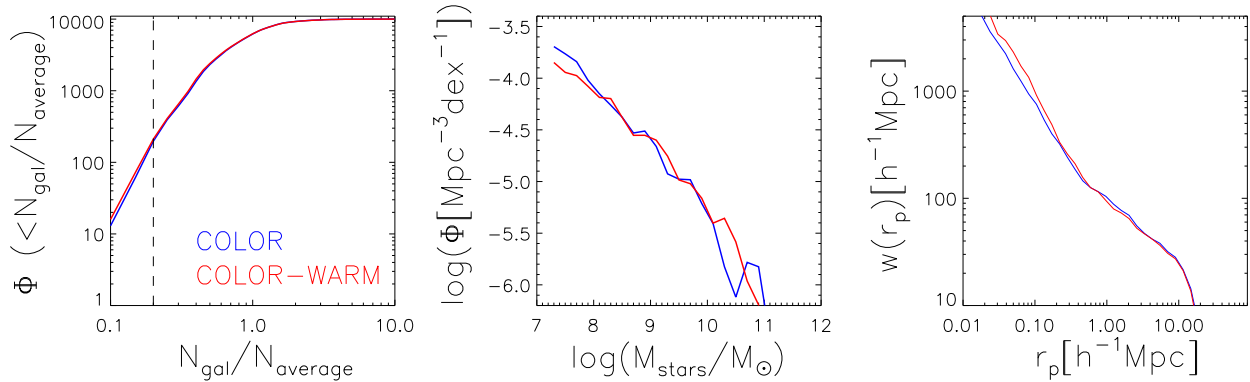}}\\%
\caption{ 
Left panel: cumulative number counts of spheres
of radius $10h^{-1}$Mpc, as a function of the ratio between galaxy number in the
sphere and the average galaxy number in a sphere of the same radius. We define
voids as those spheres with ratios smaller than $0.2$ \citep{neyrinck2008}; 
this criterion is shown by the vertical dashed line.
Middle panel: mean stellar mass function of galaxies in voids.
Right panel: two point projected correlation functions of void galaxies.
}
\label{fig:void}
\ec
\end{figure*}

Voids in the distribution of galaxies have been noticed for several decades 
\citep[e.g., ][]{joeveer1978, tully1988, peebles2001}, and are large, 
under-dense regions in the Universe. Many algorithms
define voids as underdense spheres \citep[e.g.][]{kauffmann1991, muller2000, colberg2008}. 
More sophisticated 
algorithms find voids without making any assumption about their shape \citep[e.g. 
][]{platen2007, neyrinck2008}. 
In the Local Universe, typical sizes of voids in 
the galaxy distribution vary from around 6 Mpc to more than 20 Mpc, in a 
galaxy survey like SDSS \citep[e.g.][]{ceccarelli2013}.
Voids in WDM cosmology have been investigated in the recent works of 
\citet{reed2015} who studied galaxy clustering and void volume fraction, 
and \citet{yangLF2015}, who measured the statistics and 
density profiles of voids in CDM cosmology and three WDM cosmologies
with $m_x=1.4,0.8$ and $0.4$kev.

For our purposes, we define voids in our z=0 galaxy catalogue as under-dense 
spheres of a fixed  radius $10^{-1}h$Mpc; we choose this value because it is a
typical scale of voids probed by galaxy redshift surveys \citep{ceccarelli2013}. 
We have checked that when choosing alternative radius of $5^{-1}h$Mpc or 
$15^{-1}h$Mpc, the results showing below remain similar.
For each of our simulations  we first
choose $10000$ random points in the whole simulation box as the
centers of spheres. We then compute the galaxy number density in each
sphere. All galaxies of stellar mass $>10^7h^{-1}M_{\odot}$ are 
considered. Galaxies with this lower limit stellar mass 
are typically hosted by haloes with masses of $10^{10-10.5}h^{-1}M_{\odot}$ 
(see the stellar mass -- halo mass relation in Fig.~\ref{fig:mstarmhalo}).
Such haloes contain $\gtrsim1000$ particles and are therefore well 
above the resolution limit of the simulations. 

In the left panel of Fig.~\ref{fig:void} we show the cumulative
number counts of these randomly placed spheres as a function of the ratio 
between the number of galaxies in the sphere and $N_{average}$, the average number of 
galaxies in a sphere of the
same volume over the whole simulation box. $N_{average}=246$ for COLOR,
and for COLOR-WARM $N_{average}=202$.
We define a sphere as a void when
the number of galaxies included in the sphere is less than 20~per~cent of
the average, which number is often used by the previous works studying properties of voids
\citep[e.g.][]{neyrinck2008}. This criterion is shown as a vertical dashed line
in the left panel of the figure. By this definition, there are 211
voids in the COLOR-WARM simulation, which is close to the 197 voids identified in COLOR. 

We refer to galaxies that reside in voids as void galaxies. The middle and 
right panels of Fig.~\ref{fig:void} present the
properties of void galaxies in the two models. The middle panel shows
the average stellar mass function of void galaxies, which is broadly similar
in the two models. The right panel of Fig.~\ref{fig:void} shows the two-point projected 
correlation function of void galaxies. The results are again similar in the 
two models. Our finding is consistent with the recent work of 
\citet{reed2015}, who compared galaxy clustering and void (defined as 
spheres containing zero galaxies) volume fraction between WDM and CDM 
cosmologies, and found almost no difference between the two.  

%Note that the correlation function of void galaxies shows a clear 
%break power law, with a much steeper slope at scales less than 
%$\sim 0.3h^{-1}Mpc$. This is quite different from the normally seen
%near-power law slope of correlation functions of galaxies
%\citep[e.g.,][]{li2006}. 

\subsection{Galaxy abundance in  the Milky Way and the Local Volume}

\subsubsection{The abundance of Milky Way satellites}
The CDM cosmogony gives rise to many more substructures 
in halos of mass comparable to that of the Milky Way and M31 than there are 
known satellites of those two galaxies \citep{klypin1999,moore1999,stadel2009}. 
This discrepancy is known as the missing satellite problem. The widely 
accepted solution to this problem is that some substructures fail to host 
enough star formation due to baryonic effects and/or reionization 
\citep[e.g.][]{kauffmann1993, benson2002, li2010, governato2015,sawala2016,sawala2016b}. 
Recent semi-analytic models combined with high resolution simulations
predict satellite counts consistent with recent
observations in both the CDM cosmology \citep[e.g.][]{guo2011, guo2015} and a WDM
cosmology with $m_x=3.3{\rm kev}$ \citep{bose2016}.

\begin{figure}
\bc
\hspace{-0.4cm}
\resizebox{8cm}{!}{\includegraphics{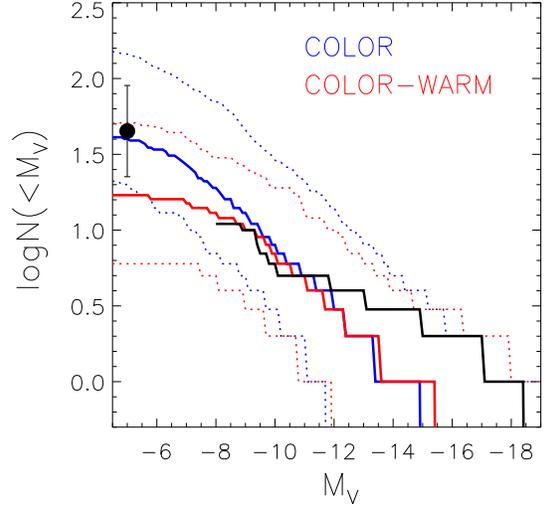}}\\%
\caption{Cumulative luminosity function for MW-analogues satellite 
galaxies. Solid color curves show the median value and dotted color
curves give 10 and 90 ~per~cent distributions of the results in the two models. 
The black line shows the result 
for the 11 classical satellites of the Milky Way, and the black dot 
is the estimation of \citet{koposov2008}, while the error bar 
is assigned to be a factor of 2 to account for the large intrinsic uncertainty. 
 }
\label{fig:satLF}
\ec
\end{figure}

Following \citet{guo2011}, we
select Milky Way-like galaxies to be disk-dominated central galaxies
with bulge-to-total mass ratio greater than $0.5$, and with a stellar mass 
between $4\times10^{10}$ and $8\times10^{10}M_{\odot}$. 
120 MW-analogues are selected in COLOR and 68 MW-analogues are selected in COLOR-WARM. 
Satellites of each MW-analogue are defined
as all galaxies within a sphere of radius $280$~kpc of the central galaxy. 
Fig.~\ref{fig:satLF} presents the cumulative luminosity
function for MW-analogue satellite galaxies in our two models.
The black line gives the result for the
11 classical satellites of the Milky Way, and the black dot is the
estimation of \citet{koposov2008}, which is based on the estimate of 45 MW 
satellites with $M_V<-5$ and $r<280$kpc from SDSS DR5. Recently more stellites
of MW have been discovered by the Dark Energy Survey 
\citep[DES;][]{bechtol2015, drlica2015,koposov2015},
the VST-Atlas\citep{torrealba2016} and Pan-STARRS $3\pi$ surveys\citep{laevens2015a,laevens2015b}.
Fig.11 of \citet{bose2016} has included these latest observed satellites, which 
is consistent with the data shown in our Fig.~\ref{fig:satLF},
with more data points and extending to fainter luminosity.
However, the newly found MW satellites have anisotropic distribution,
and the completeness is quite uncertain \citep[for detail discussion see][]{lovell2015}. 
We therefore choose to show here the more conservative observational data as reference.

In Fig.~\ref{fig:satLF}, regions between dotted lines indicate the 10th and 
90th percentile distribution of the two model results, and both cover the observation.
The results from the two models are quite similar
for magnitudes $M_{\mathrm{V}}$ brighter than $-10$. At fainter magnitudes,
MW-analogues in COLOR host more satellites than those in
COLOR-WARM. The difference in the median becomes apparent for
satellites with absolute magnitude fainter than $-8$. At $M_V=-6$, the
difference in the median is as large as $0.4$ dex. 
The difference between CDM and WDM cosmologies in our models is larger 
than that found in \citet[][Fig.10]{bose2016}. This is mainly due to 
the fact that the WDM particle we adopt has a longer free-streaming length 
than the $3.3${\rm kev} particle used in their paper, which greatly reduces 
the number of low-mass dark matter haloes.
Other than that, effects caused by different simulations used like cosmic variance
may also play a role.

\subsubsection{The Local Void}
Full sky galaxy surveys in the local Universe have revealed that 
about a third of the Local Volume contains only 3 galaxies \citep{peebles2001,
peebles2010}. 
As noted above, these surveys are about 80 ~per~cent complete to 
$m_{\mathrm{B}} = 17.5$ \citep{karachentsev2014}. 
The emptiness of this region, known as Local Void, has been
claimed as a potential challenge to the standard
$\Lambda$CDM cosmogony \citep{peebles2010}. \citet{xie2014}
studied the probability of finding a region like the Local Void in
a simulated $\Lambda$CDM galaxy catalogue. They found a
probability as high as 14~per~cent, which suggests that the emptiness
of the Local Void may not be in conflict with $\Lambda$CDM.
As the Local Volume containing the Local Void provides a laboratory
for observing the faintest galaxies in the Universe, 
we study statistics of ``Local Volumes'' and ``Local Voids'' in both
our WDM and CDM  catalogues to investigate if any differences can be
measured. 

Following the method and definitions of \citet{xie2014}, ``Local Volumes'' 
in the simulations are defined as regions within a radial range of [$1,8$Mpc]
around simulated MW-analogues, which are selected to be central 
disk galaxies with bulge to-total stellar mass ratio less than $0.5$, and 
with B-band magnitude within the range $[-21.5,-19.5]${\footnote{Here we 
choose the criterion of B-band magnitude rather than stellar mass to select 
MW-like galaxies, to get more simulated ``Local volumes'' 
for statistical study. There are too few ``Local volumes'' if MW-analogues
are selected by stellar mass, due to the lack of M* galaxies in the stellar 
mass functions in our models as seen in Fig. 1.}. Since our MW and M31 are
separated by a distance of 0.77Mpc,
``Local volumes'' are further selected to have a
companion giant galaxy with a stellar mass similar to that of the M31,
with stellar mass in the range [$0.5\times M_{MW}$, $2\times M_{MW}$],
within 1~Mpc from the ‘Milky Way’{\footnote{We have checked that our 
conclusion in this subsection will not be affected if this distance of 1 ~Mpc is
changed to 2 ~Mpc.}}. In addition, systems with clusters
(haloes more massive than $10^{14}M_{\odot}$) close by ($<10$ Mpc) are excluded,
to mimic the relative isolated environment of the Milky Way, where
the Virgo cluster is about 16.5 Mpc away \citep{mei2007}. 
We thus determine that there are 105 thus-defined local volumes in COLOR 
and 100 in COLOR-WARM. 

Fig.~\ref{fig:localvolume} presents a comparison of the luminosity functions
of galaxies in the simulated ``Local Volumes''. 
As reference, black line with error bars gives the observed luminosity function based
on the data provided by \citet{karachentsev2014}, including all galaxies within 
1-8Mpc to be consistent with the simulated ``Local Volumes''. Note that the
completeness of the observed galaxies should be different at different magnitues, 
with brighter galaxies being more completely observed. However, the completeness 
is still  uncertain in observation and therefore not corrected for the 
observation result shown.

Fig.~\ref{fig:localvolume} shows that the bright end of the luminosity
function in the two models are similar. Observation at the bright end is higher
than the median values of both models. 
At the faint end, COLOR-WARM produces lower amplitude up to a factor 
of 2,  and thus achieves a better match to the observation than does COLOR. 
Apart from the luminosity function of galaxies in the Local Volume, \citet{klypin2015}
and \citet{schneider2016} studied the HI velocity function 
of galaxies in the Local Volume in both CDM and WDM models. 
\citet{schneider2016} found that the discrepancy between observation and CDM model
disappears in warm and mixed (warm plus cold) dark matter models.

\begin{figure}
\bc
\hspace{-0.4cm}
\resizebox{8cm}{!}{\includegraphics{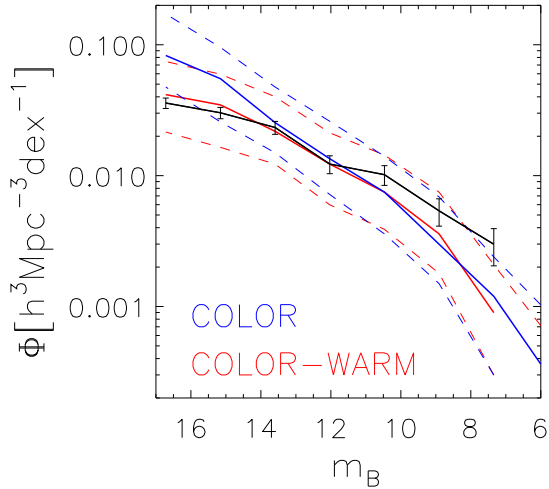}}\\%
\caption{ 
The apparent B band magnitude luminosity function of galaxies in 
simulated local volumes. The observed luminosity function calculated
based on the data provided by
\citet{karachentsev2014} is shown as a black line. Simulated local 
volumes are shown by red (COLOR-WARM) and blue (COLOR) lines, with solid 
lines showing the median value, and dashed lines show the 68 percentile envelope 
of the distribution.
}
\label{fig:localvolume}
\ec
\end{figure}

Note that galaxies shown in Fig.~\ref{fig:localvolume} can have stellar 
masses as low as around $10^6M_{\odot}$, while for the stellar mass functions
shown in Fig.~\ref{fig:SMF}, only galaxies more massive than $10^8M_{\odot}$  
are presented, and in Fig.~\ref{fig:void}, galaxies more massive than 
$10^7M_{\odot}$ are included. If we also plot the SMF down to $10^6M_{\odot}$, there 
are more galaxies in the COLOR box as a whole as well (see Fig.~\ref{fig:LFall}), which is 
not in conflict with the results seen here. On the other hand, 
Fig.~\ref{fig:LFall} indicates that SMFs become shallower below a few times
$10^6M_{\odot}$, which marks the resolution limit of the two models. Therefore,
results shown in Fig.~\ref{fig:localvolume} are not quite complete. However,
with the trend that the two models have larger difference towards lower 
masses, the results in Fig.~\ref{fig:localvolume} should remain similar
if a higher resolution simulation is used.  

For our next step, we follow \citet{xie2014} in identifying simulated 
``Local Voids'' as the most empty, truncated, cone-shaped region with solid 
angle $\pi$. Fig.~\ref{fig:localvoid} 
shows the cumulative fraction of the 
simulated ``Local Voids'' as a function of the number of galaxies enclosed in the 
void. Simulated  ``Local Voids'' in COLOR-WARM are overall emptier than the ones in
COLOR. The observed Local Void around the real Milky Way contains fewer than 
5 galaxies (when Poisson noise is taken into account), indicated by 
the vertical dashed line in Fig.~\ref{fig:localvoid}. Statistically, $5.7$ 
~per~cent of simulated Local Voids are as empty as the observed Local Void 
in COLOR \footnote{Note that this probability is somewhat different from the $14$ ~per~cent 
result given by \citet{xie2014}. The difference is expected because of
the different simulations used, the different galaxy formation models applied, 
and also due to cosmic variance. }.
The number of such voids in COLOR-WARM is much higher, up to approximately $34$~per~cent. 
Therefore, in our models the probability of hosting the Local Void is more than 5 
times higher in this WDM model than in the CDM cosmology.

\begin{figure}
\bc
\hspace{-0.4cm}
\resizebox{7.5cm}{!}{\includegraphics{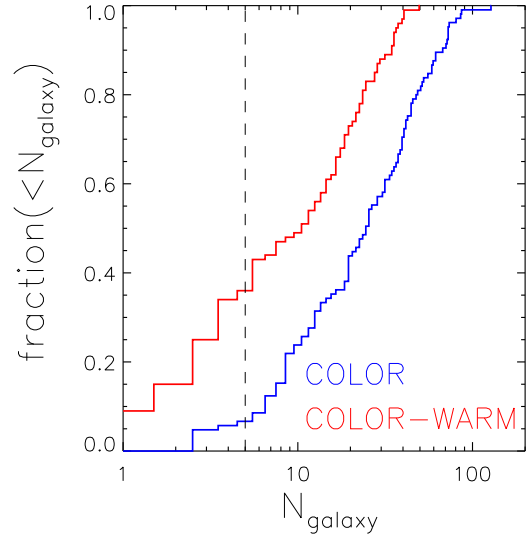}}\\%
\caption{ 
The cumulative fraction of simulated Local Void samples as a 
function of the number of galaxies enclosed in the void. The vertical 
dashed line indicates the number of 5 galaxies observed in the real Local Void.
}
\label{fig:localvoid}
\ec
\end{figure}

%%%%%%%%%%%%%%%%%%%%%%%%%%%%%%%%%%%%%%%%%
\section{conclusion and discussions}
\label{sec:con}

This work explores how properties of the galaxy population change under
different assumptions about the mass of dark matter particles, going from a CDM
to a WDM cosmogony. We have chosen a warm dark matter (WDM) particle with
a linear power-spectrum consistent with a $1.5$~{\rm kev}
thermal relic, which is warmer than the latest Lyman-$\alpha$ forest constraint
allows, to maximize the differences that arise. The two simulations we use,
COLOR and COLOR-WARM, have sufficient resolution and volume to explore this
problem with unprecedented statistics and precision.

We combine the two simulations with the {\sc GALFORM} semi-analytic model of
galaxy formation and evolution. We apply the  {\sc GALFORM} model of
\citet{gonzalez2014} in our CDM simulation. In our WDM simulation, we retune
the parameters of this model to bring the galaxy stellar mass function at $z=0$
into agreement with that of the CDM simulation in the mass range
$10^9<M_{stars}/(h^{-1}M_{\odot})<10^{11}$.

We have compared various properties of the model galaxy population to $z=10$.
Many of the properties we examine are indistinguishable between the two dark
matter models, including the spatial clustering of galaxies (Fig.2), and the
statistics of the galaxy population in under-dense regions (Fig.6).  However,
differences become more obvious at higher redshifts, with massive galaxies
forming more efficiently in the WDM model. This is indicated by results showing
that the high mass end of the stellar mass functions in the WDM model are up to
about 3 and 9 times higher than in CDM at $z=6$ and $z=10$ respectively
(Fig.3). Moreover, at $z>3$ the star formation rate density in the WDM model is
up to 2 times higher (Fig.4). At $z=0$, a transition is evident at stellar mass
of around $10^{10}M_{\odot}$ (Fig.5). Galaxies less massive than
$10^{10}M_{\odot}$ assemble both stellar mass and halo mass later in WDM
compared to a CDM cosmology. 
The difference is greatest for low mass galaxies
(stellar mass $10^{8.5}M_{\odot}$), which is more than 100 times for the
mean stellar mass and more than 30 times for the mean halo mass at $z>9$.
Galaxies more massive than $10^{10}M_{\odot}$ assemble their stellar masses
somewhat earlier in the WDM model, with differences in stellar mass up to a
factor of 3 for redshifts between $\sim 3$ and $10$.
The difference between CDM and WDM models at $z>7$ is also seen 
in \citet{bose2016}, with a rest mass of $3.3kev$ in WDM model adopted, but
the difference is indistinguishable by current observation. 

In the local universe, the most pronounced differences between the two dark
matter models are found in the counts of galaxies in analogues of the Local
Volume and the probability of observing a region as empty as the Local
Void (section 4.2.2). The simulated ``Local Volumes'' have lower number
density by up to a factor of 2 at the faint end of the luminosity function in
the WDM model (Fig.8).  The probability of a region as empty as the
observed Local Void is as high as $34$ ~per~cent in the WDM model, 5 times
higher than that in the CDM model (Fig.9). A more complete Local Volume galaxy
census \citep[e.g. the Dark Energy Spectroscopic Instrument's Bright Galaxy
Survey, DESI-BGS, ][]{DESI2016} should therefore provide a way to
constrain the nature of dark matter.

%%%%%%%%%%%%%%%%%%%%%%%%%%%%%%%%%%%%%%%%
\section*{Acknowledgments}
We acknowledge Ran Li, Hong Guo for the helpful discussions. LW acknowledges
support from the NSFC grants program (No.
11573031, No. 11133003). VGP Acknowledges past 
support from the European Research Council Starting Grant (DEGAS-259586). 
WAH is supported by the European Research Council grant through
646702 (CosTesGrav) and the Polish National Science Center under 
contract \#UMO-2012/07/D/ST9/02785.

The work was carried out at National
Supercomputer Center in Tianjin, and the calculations were performed on
TianHe-1（A）.

This work used the DiRAC Data Centric system at Durham University, operated 
by the Institute for Computational Cosmology on behalf of the STFC DiRAC HPC 
Facility (www.dirac.ac.uk). This equipment was funded by BIS National 
E-infrastructure capital grant ST/K00042X/1, STFC capital grants ST/H008519/1 
and ST/K00087X/1, STFC DiRAC Operations grant ST/K003267/1 and Durham 
University. DiRAC is part of the National E-Infrastructure.

This work is part of the D-ITP consortium, a programme of the Netherlands 
Organization for Scientific Research (NWO) that is funded by the Dutch 
Ministry of Education, Culture and Science (OCW).

%\bsp
\label{lastpage}

\bibliographystyle{mn2e}
\bibliography{wdm}

%%%%%%%%%%%%%%%%%%%%%%%%%%%%%%%%%%%%%%%%%%%%%%%%%%%%%%%%%%%%%%%%%%%%%%%%%%%%

\appendix
 \section{Halo mass functions in simulations}

\begin{figure}
\bc
\hspace{-0.4cm}
\resizebox{7.5cm}{!}{\includegraphics{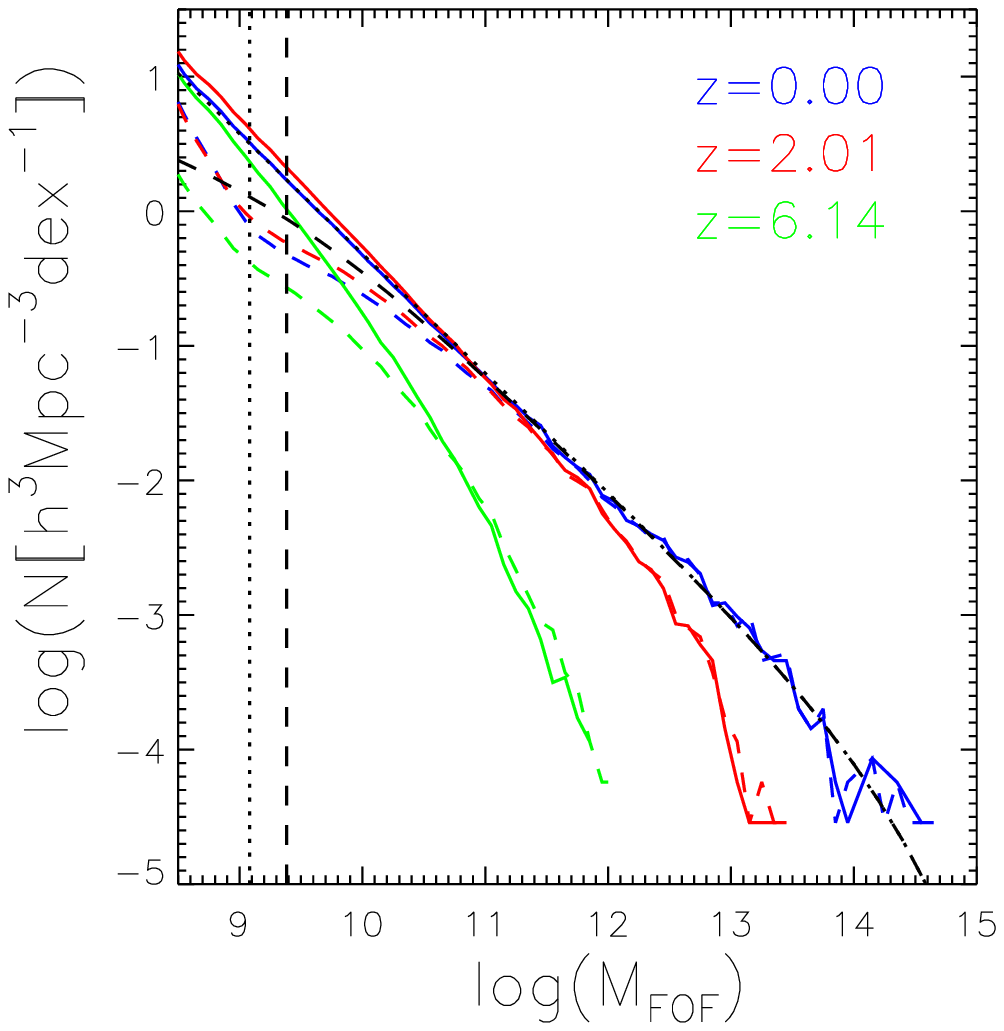}}\\%
\caption{ FOF group mass (total particle mass) function at three redshifts 
in the simulations. Solid lines denote the COLOR simulation and dashed lines COLOR-WARM. 
The vertical dashed and dotted lines are the limiting 
masses below which spurious halos form in COLOR-WARM, given by \citet{wangjie2007} and 
\citet{lovell2014}.
}
\label{fig:haloMF}
\ec
\end{figure}

Fig.~\ref{fig:haloMF} shows the FOF halo mass functions at different 
redshifts in the COLOR and COLOR-WARM simulations as described 
in Section 2.1. WDM halo mass functions flattens towards the low mass end compared 
to those in CDM. As first indicated by  \citet{wangjie2007}, spurious haloes 
form due to a numerical discreetness effect in warm and hot dark matter simulations.
The mass limit of dark matter haloes below which spurious haloes form depends
on the N-body particle mass of the simulation and the WDM particle mass assumed.
In COLOR-WARM, the N-body mass resolution is $m_{\rm dm}=6.20\times10^{6} h^{-1}M_{\odot}$, and the
WDM particle mass is $m_{x}=1.5{\rm kev}$. This corresponds to a mass limit of 
$M_{lim}=2.42\times10^{9}\,h^{-1}{\rm M}_{\odot}$ and $M_{min}=0.5M_{lim}$
given by equations derived by \citet{wangjie2007} and \citet{lovell2014} respectively. 
These limits are indicated
by a dashed and a dotted vertical lines in Fig.~\ref{fig:haloMF}. Below these mass limits,
the halo mass functions steepen significantly due to the emergence of spurious haloes. 

\section{Box and resolution effects}\label{app:resolution}

\begin{figure*}
\bc
\hspace{-0.4cm}
\resizebox{14.cm}{!}{\includegraphics{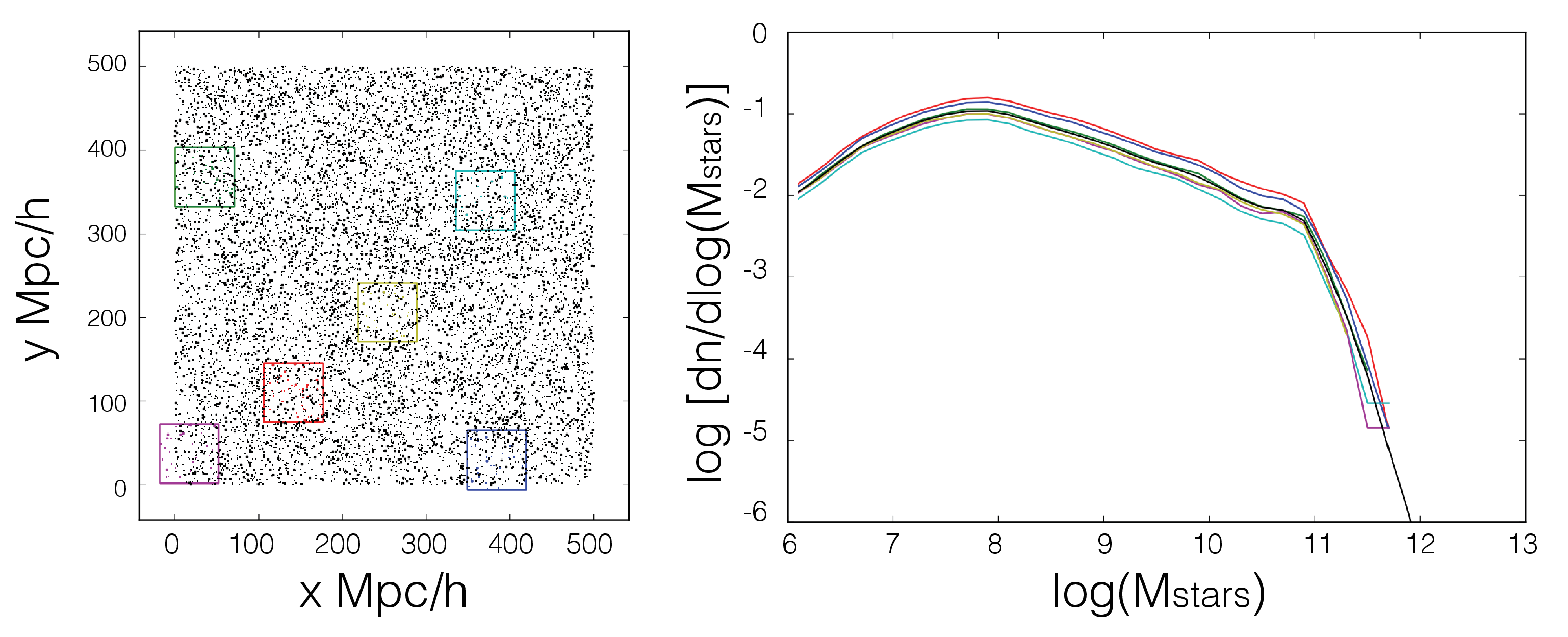}}\\%
\caption{Left panel: six random COLOR-sized regions (small coloured boxes) are extracted from
MR-W7 (black dots denote the whole box). Right panel: the galaxy stellar mass functions 
of the selected COLOR-sized regions (line colour corresponds to the box colour in the
left panel) combined with GP14 model. The black line is the stellar mass function of MR-W7
combined with the GP14 model. 
 }
\label{fig:cosmic_variance}
\ec
\end{figure*}

In this study, the simulation boxes we use are much smaller than the 
MS-W7 simulation that the GP14 model was calibrated with. In addition, the mass 
resolution of our simulations is about 100 times higher than that of MS-W7. 
Fig.~\ref{fig:cosmic_variance} shows the galaxy SMFs at $z=0$ of six 
COLOR-sized regions extracted from the MR-W7 simulation. We find that variations of up to 
0.3 dex in galaxy abundance are expected due to cosmic variance.
 
Fig.~\ref{fig:resolution} shows the SMFs of the GP14 model combined with MR-W7 and 
COLOR, while the dotted line is the SMF when COLOR trees have all the branches 
below the resolution of MR-W7 removed. Comparing the blue solid and blue 
dotted lines, we find that the effect of different mass resolutions on SMF is 
very small.
The difference between the published GP14 model and that run using merger trees 
from the COLOR simulation is mainly due to cosmic variance.

\begin{figure}
\bc
\hspace{-0.4cm}
\resizebox{7.5cm}{!}{\includegraphics{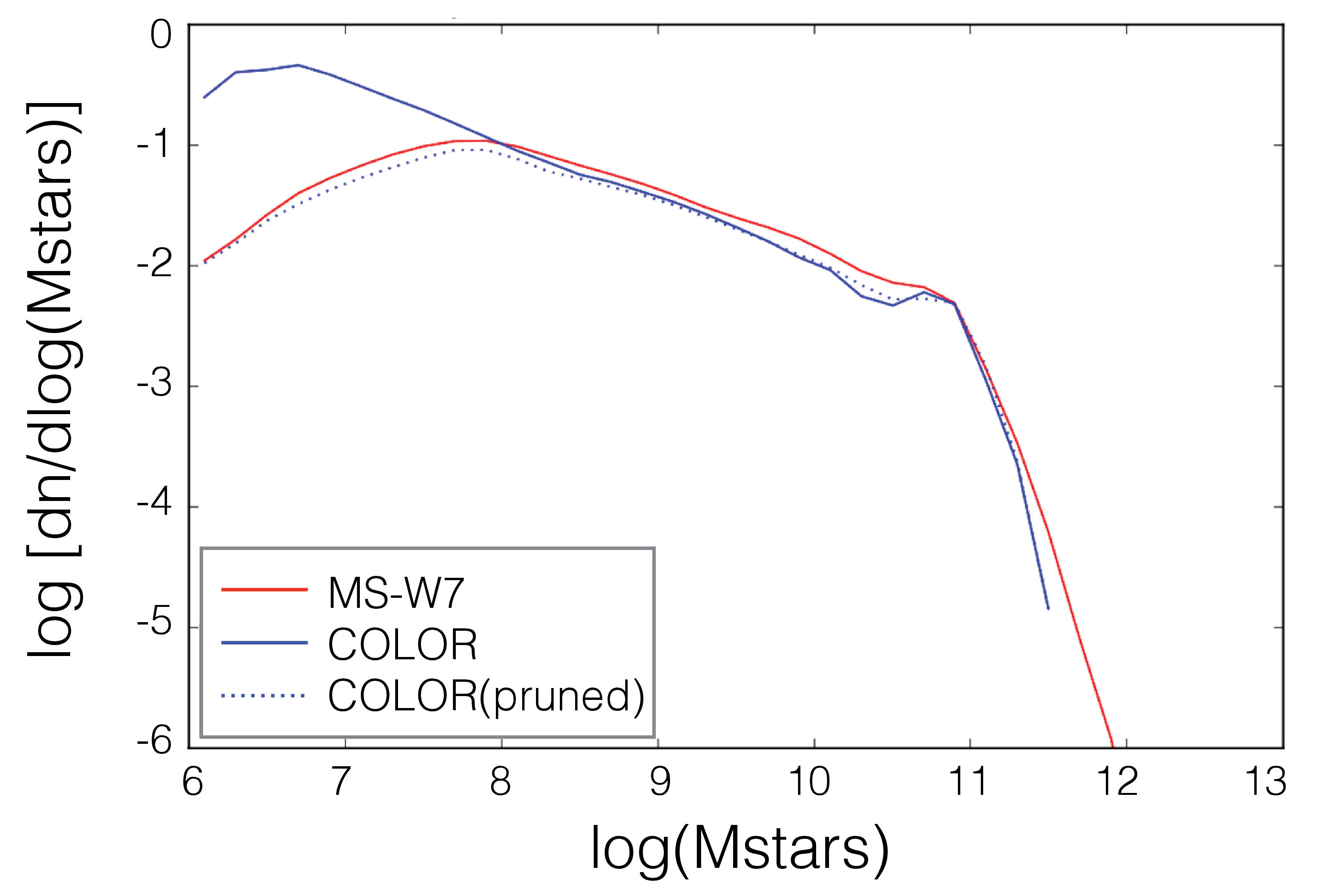}}\\%
\caption{Stellar mass functions in the COLOR (blue line) and MR-W7 (red line), 
combined with GP14 model. The dotted line shows the result when 
the COLOR trees have all the branches below the resolution of 
MR-W7 removed.}
\label{fig:resolution}
\ec
\end{figure}

\section{The Re-GP14 model}\label{app:regp14}

 \begin{table}
 \caption{Changes of parameters from the fiducial GP14 model to the 
retuned GP14 model, in order to 
match the galaxy stellar mass function at $z=0$ for $10^9<M_{stars}/(h^{-1}M_{\odot})<10^{11}$ of COLOR-WARM to that of COLOR.}
\begin{center}
 \begin{tabular}{cccccccc} \hline
  Parameter  &   GP14 & Re-GP14 \\ \hline
    $\alpha_{hot}$  &   3.2  &  2.5  \\ 
    $V_{hot,disk}$  &   425  & 575  \\ 
    $\alpha_{cool}$ &  0.6  & 0.55  \\ \hline
 \end{tabular}
\end{center}
 \end{table}

For this study we consider two galaxy formation models, one for COLOR (GP14) 
and a second for COLOR-WARM (Re-GP14), constructed to have very similar galaxy stellar mass
functions at z=0 for $10^9<M_{stars}/(h^{-1}M_{\odot})<10^{11}$.
Changing the parameters related to feedback is sufficient for the
Re-GP14 model in COLOR-WARM to reproduce the stellar mass function from COLOR combined with GP14. 
The detailed changes of parameters from GP14 to Re-GP14 are listed in Table.~C1. 
Both $\alpha_{hot}$ and $V_{hot,disk}$ are related to supernovae 
feedback, in that they determine the relation between the ejected cold gas mass and 
circular velocity ${V}_{circ}$. $\alpha_{hot}$ is the exponent of the relation 
(equation 1 of GP14):
\begin{displaymath}
{\dot{M}_{reheated}}\propto({V}_{circ}/{V}_{hot,disk})^{-{\alpha}_{hot}}
\end{displaymath}
Decreasing $\alpha_{hot}$ leads to lower feedback and ejected mass in low circular velocity 
galaxies and higher feedback in high circular velocity galaxies. 
$V_{hot,disk}$ is the normalization of the relation for disks. The efficiency of supernovae 
feedback increases with larger $V_{hot,disk}$. The normalization for bulges 
has not been changed. 

\begin{figure}
\bc
\hspace{-0.4cm}
\resizebox{7.5cm}{!}{\includegraphics{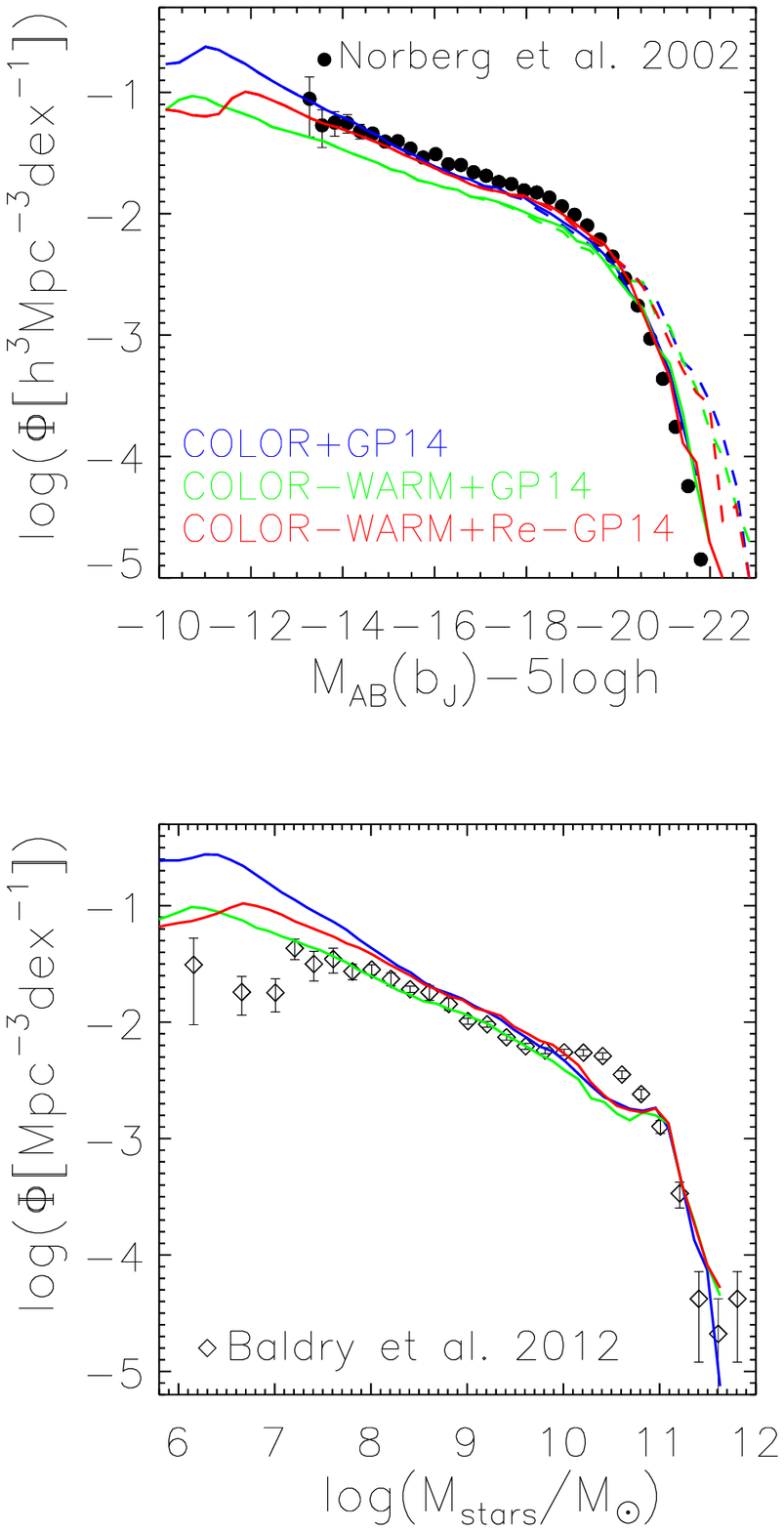}}\\%
\caption{ $b_J$ band luminosity functions and stellar mass functions of 
galaxies simulated. The symbols and lines in the panels are the same 
as in Fig. 1, whilst the additional green lines are 
the results of COLOR-WARM simulation combined with the original GP14 model. 
Note that the x-axis extends to lower stellar masses than in Fig.~1.
 }
\label{fig:LFall}
\ec
\end{figure}

Overall, supernovae feedback is weaker in low mass galaxies, which 
leaves more cold gas available for star formation and therefore more 
efficient star formation in low mass galaxies, and results in a larger 
amplitude for the luminosity and stellar mass functions. In massive 
galaxies, supernovae feedback is stronger.

$\alpha_{cool}$ controls when a halo can be affected 
by AGN feedback \citep[see equation 12 in ][]{lacey2015}. In Re-GP14, 
smaller values of $\alpha_{cool}$ correspond to
AGN feedback affecting fewer galaxies, which in turn alters the bright-end slope of the luminosity and 
stellar mass functions. 
 
\begin{figure}
\bc
\hspace{-0.4cm}
\resizebox{8cm}{!}{\includegraphics{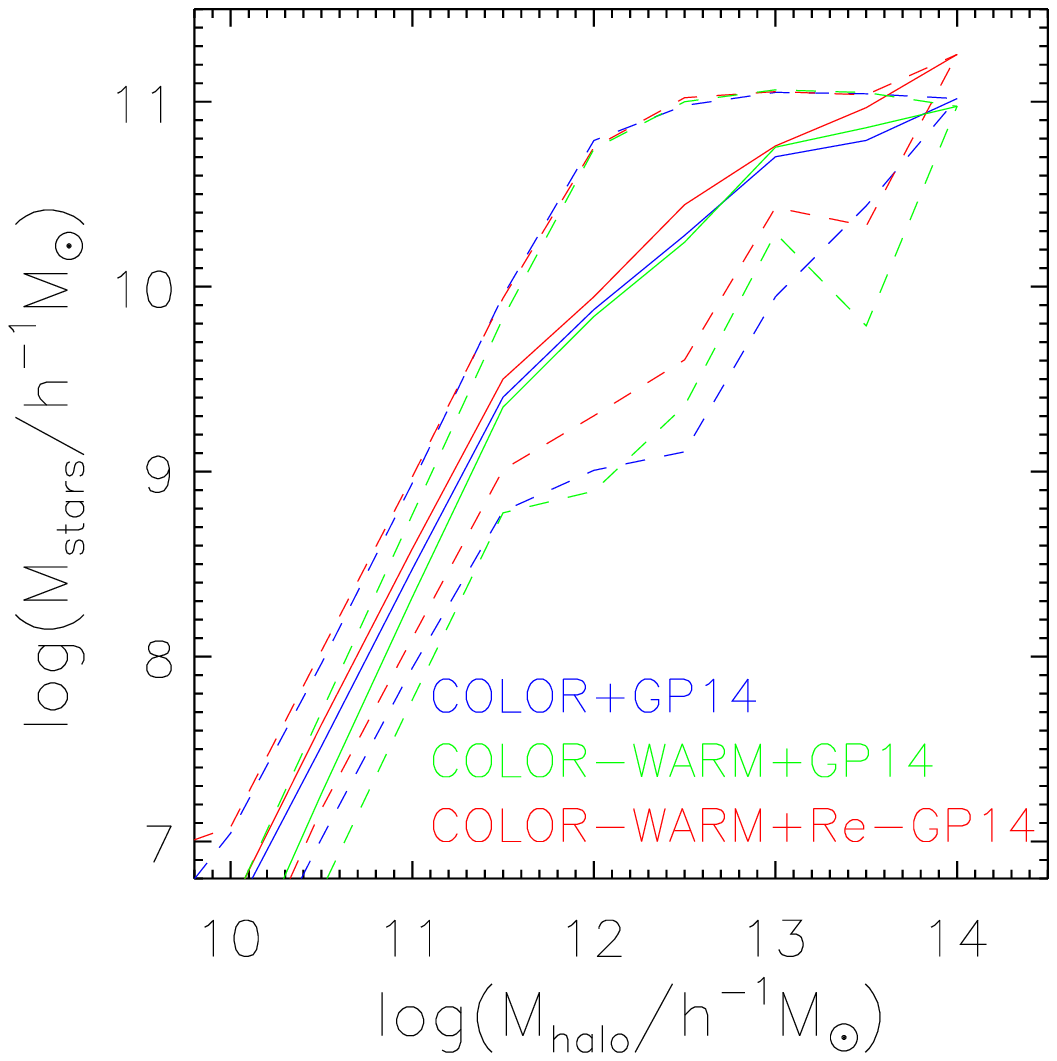}}\\%
\caption{ Median stellar mass - halo mass relations of central galaxies in the 
three models shown in the legend. Dashed lines indicate the region containing 90~per~cent of the distribution for 
each relation.
 }
\label{fig:mstarmhalo}
\ec
\end{figure}

Fig.~\ref{fig:LFall} shows the $b_J$ band luminosity functions and 
stellar mass functions of galaxies in three catalogues at z=0. Green lines 
show results from COLOR-WARM simulation combined with the original GP14 model 
without tuning of parameters. With the same semi-analytic models, COLOR-WARM 
produces fewer small galaxies because there are fewer low mass haloes than in the 
COLOR simulation. Compared with the luminosity function results given by 
\citet{bose2016}, where the thermal mass of the WDM particle is assumed
to be $3.3${\rm kev} for the WDM simulation, the difference between WDM and CDM is larger
in our models, since the WDM particle mass assumed in our model,  $1.5${\rm kev}, 
is warmer. The overall main effect of changing parameters from GP14 to 
Re-GP14 is that the number of low mass/luminosity galaxies is larger, bringing
galaxies in COLOR-WARM closer to the COLOR ones. The remaining difference
on the faint/low mass end at stellar masses less than $\sim 10^{8}M_{\odot}$ is 
mainly due to the fact that there are fewer small haloes in COLOR-WARM, 
and thus not from the retuning of the galaxy formation model. 

Fig.~\ref{fig:mstarmhalo} shows the stellar mass -- halo mass relations in the
three models. With the same GP14 model, WDM produces lower stellar mass  
galaxies at fixed halo mass than CDM
for haloes of mass  $<10^{11.5}h^{-1}M_{\odot}$. When combined with the
Re-GP14 model, with weaker feedback effect, the stellar mass in COLOR-WARM becomes 
overall larger.

%%%%%%%%%%%%%%%%%%%%%%%%%%%%%%%%%%%%%%%%%%%%%%%%%%%%%%%%%%%%%%%%%%%%%%%%%%%%
\end{document}